\documentclass[amsmath,amssymb,aps,prl,superscriptaddress,reprint,nofootinbib]{revtex4-2}

	\usepackage{graphicx}
	\usepackage{soul}
	\usepackage[colorlinks=true,citecolor=blue,linkcolor=magenta]{hyperref}
	\usepackage[usenames]{color}   
	\usepackage{amsfonts}
	\usepackage{color}
	\usepackage{booktabs}
	\usepackage{multirow}
	\usepackage{float}
    \usepackage[english]{babel}
	\usepackage{textcomp}
	\usepackage{upgreek}
    \usepackage{physics}
    \usepackage{siunitx}
    \usepackage{xr-hyper}
    \usepackage{hyperref}
    \usepackage{ulem}
    \makeatletter
    \renewcommand\@make@capt@title[2]{%
      \@ifx@empty\float@link{\@firstofone}{\expandafter\href\expandafter{\float@link}}%
      {\textbf{#1}}\@caption@fignum@sep#2%
      }%
    \makeatother

\begin{document}

\title{Fast Nondestructive Readout for High-Clock-Rate Atom Array Quantum Processor}



\author{Xu-Zhao-Qiu Zeng}
\thanks{These authors contributed equally to this work.}

\author{Chang You}
\thanks{These authors contributed equally to this work.}

\author{Qing-Wei Wang}

\author{Zi-Feng Li}

\author{Yi Ji}

\author{Dong An}

\author{Chao Yu}

\author{Jia-Rui Liu}

\author{Zi-Mo He}

\author{Jia-Rui Gu}

\author{Yuhao Mei}

\author{Hao-Wen Cheng}

\author{Yu-Chen Zhang}

\author{Rui Lin}

\author{Zhan Wu}

\author{Jun Rui}

\author{Jun Zhang}

\author{Ming-Cheng Chen}

\author{Yu-Hao Deng}
\email[Correspondence: ]{dengyh@ustc.edu.cn}

\author{Chao-Yang Lu}
\email[Correspondence: ]{cylu@ustc.edu.cn}

\author{Jian-Wei Pan}
\email[Correspondence: ]{pan@ustc.edu.cn}
\affiliation{Hefei National Research Center for Physical Sciences at the Microscale and Department of Modern Physics, New Cornerstone Science Laboratory, University of Science and Technology of China, Hefei 230026, China}
\affiliation{Shanghai Research Center for Quantum Science and CAS Center for Excellence in Quantum Information and Quantum Physics, University of Science and Technology of China, Shanghai 201315, China}
\affiliation{Hefei National Laboratory, University of Science and Technology of China, Hefei 230088, China}

\begin{abstract}

Neutral-atom arrays have rapidly advanced to support thousands of qubits and execute high-fidelity logical operations. However, these processors remain severely throttled by their slowest fundamental operation: nondestructive qubit measurement, which requires milliseconds and fundamentally limits the system's clock rate. 
This bottleneck arises from both an inherent photon-budget dilemma---sufficient fluorescence for reliable state discrimination must be collected without excessive heating or loss---and frame-based imaging, which imposes one common exposure and decision latency on intrinsically independent, site-local measurements.
Here, we overcome these limitations with a fast, nondestructive readout architecture based on real-time, site-resolved adaptive protection.
By integrating continuous photon counting with a dynamic feedforward framework, we decode qubit states with sub-microsecond latency and instantly shield atoms from redundant scattering.
Demonstrated in parallel across a 100-qubit reconfigurable atom array, with adaptive protection on a 25-site subarray, 
this dynamic decision protocol reduces the average probe time to just $15\ \mu\text{s}$.
Model-free benchmarking yields a discrimination infidelity of $4.1 \times 10^{-5}$ and an atom loss of $2.1 \times 10^{-4}$, simultaneously setting new performance records for atom arrays. Exploiting this capability, we operate repeated quantum circuits at an unprecedented 1.7 kHz clock rate with atoms reused over 120 consecutive rounds—nearly sevenfold higher than the previous record—and enter the sub-millisecond cycle regime for the first time. By removing nondestructive readout as the dominant cycle-time bottleneck, this work unlocks high-clock-rate mid-circuit syndrome extraction, paving the way for high-throughput, fault-tolerant quantum computation.

\end{abstract}

\maketitle

Fault-tolerant quantum computation is measurement-intensive: syndrome extraction and non-Clifford gates require repeated mid-circuit readout interleaved throughout the logic circuit \cite{terhal2015quantummemory, bravyi2005universaldistillation, gidney2024cultivation}. Neutral-atom arrays have assembled most of the required elements to a remarkable level---thousands of highly coherent qubits \cite{largearray-manetsch2025nature, largearray-chiu2025nature}, sub-microsecond entangling gate with fidelity beyond error-correction thresholds \cite{jaksch2000fastRydberg,gate-evered2023nature,tsai2025benchmarking}, and key ingredients of quantum error correction \cite{transport-bluvstein2022nature, lukin-bluvstein2024nature, camera-bluvstein2026nature}. Yet nondestructive, state-resolved readout remains at the millisecond scale \cite{camera-graham2023physrevx,camera-kwon2017physrevlett,camera-martinez-dorantes2017physrevlett,camera-bluvstein2026nature} — orders of magnitude slower than the entangling gate. This slow readout dictates the clock rate of the entire processor and causes accumulated idling errors, eroding the platform's hard-won scaling advantages \cite{zhou2025nature-transversal, saffman2025connectivitycost}.

\begin{figure*}[ht]
	\centering
	\includegraphics[width=1\textwidth]{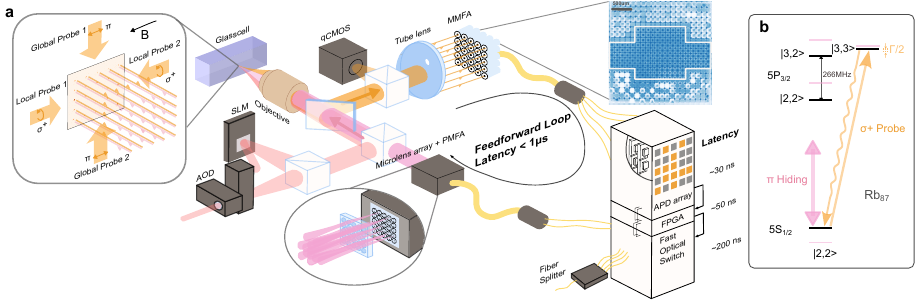}
	\caption{\textbf{FNDR architecture based on fibre-array-coupled photon counting and adaptive hiding.} 
	\textbf{a,} 
    A \(10 \times 10\) atom array is trapped by a spatial light modulator (SLM)-generated static tweezer array and dynamically rearranged by crossed acousto-optic deflectors (AODs). Fluorescence is collected by high-NA objectives and separated into a camera arm and a photon-counting arm by a PBS. The \(\pi\)-polarized channel is imaged on a qCMOS camera for wide-field detection, whereas the \(\sigma^+\)-cycling readout channel is coupled into a multimode fibre array (MMFA) for site-resolved detection by APDs.
    An FPGA processes the APD click in real-time using an adaptive strategy.
    When a site's state is identified with target confidence, the FPGA triggers a high-speed optical switch array, which directs an individual-addressing hiding beam (pink) through a polarization-maintaining fibre array (PMFA) and microlens array to suppress further cycling transitions. The closed-loop latency is \(<1~\mu\mathrm{s}\). Only a \(5 \times 5\) region is drawn, and the opposite-side objective and MMFA are omitted for clarity.
    \textbf{b,} Level scheme for nondestructive state readout. The \(\sigma^{+}\)-probe addresses the \(\ket{2,2}\rightarrow\ket{3,3}\) cycling transition, red-detuned by \(\Gamma/2\simeq3~\mathrm{MHz}\), with saturation parameter \(s_0 \approx 5\). The \(\pi\)-polarized hiding beam suppresses further photon scattering after detection while minimizing depumping.
    } 
	\label{fig1}
\end{figure*}

This severe mismatch stems from the photon-budget dilemma: fluorescence readout must collect enough photons for reliable state discrimination, yet the ensuing redundant scattering causes recoil heating, off-resonant depumping, and atom loss.
Conventional camera readout compounds this physical constraint with an architectural one: unlike quantum-gas microscopy, where submicron
spatial resolution is essential \cite{bakr2009quantumgasmicro}, atoms in a tweezer array are trapped at known, well-separated sites, so the task is not to reconstruct an image but to read out many independent, spatially fixed binary channels. 
Frame-based imaging nevertheless forces all sites to share a common exposure and defers all decisions until the frame is complete, causing most atoms to scatter photons long after sufficient information has been acquired.
Single-photon avalanche photodiodes and optical cavities mitigate the photon-budget tradeoff through time-resolved detection and enhanced photon collection, respectively \cite{apd-gibbons2011physrevlett,apd-fuhrmanek2011physrevlett,shea2020submillisecond,apd-chow2023physreva,bochmann2010losslesscavity, gehr2010cavity, fiberandcavity-grinkemeyer2025science, wang2025ultrafastcavity, deist2022midcavity,hu2025sitecavity,camera-lee2026rapidcavity}. However, previous free-space single-photon-counting demonstrations have been restricted to individual atoms, whereas cavity-enhanced approaches require specialized optical hardware that has not yet been integrated with large, reconfigurable processors.

To break this limit, we introduce a fast nondestructive readout (FNDR) architecture that converts array readout from a fixed camera exposure into a real-time, site-local adaptive decision process. By coupling fluorescence from the atom array into independent photon-counting channels, an FPGA classifies each atom's state on the fly. The instant an atom is identified as bright, a per-site hiding beam dynamically decouples it from the cycling transition. This sub-microsecond feedforward loop truncates unnecessary scattering, simultaneously resolving the speed-survival trade-off. Crucially, the FNDR integrates directly into standard neutral-atom quantum processors, enabling high-clock-rate repeated circuits with atom reuse.

\vspace{1em}
\noindent\textbf{The FNDR Architecture and Adaptive Strategy}

\noindent The FNDR architecture is implemented on a two-dimensional tweezer-atom-array platform designed for mid-circuit operation.
A \(10\times10\) single-atom array with \(5~\mu\mathrm{m}\) spacing is prepared in SLM-generated optical tweezers and cooled to \(14~\mu\mathrm{K}\), and operated under a \(10~\mathrm{G}\) bias field (Fig.~\ref{fig1}(a)).
Fast, nondestructive readout uses a pair of alternately pulsed \(\sigma^+\)-polarized probe beams \cite{destructive-su2025natcommun, yb-muzi2025microsecond}, red-detuned by \(\Gamma/2\simeq3~\mathrm{MHz}\) from the free-space \(\ket{F=2,m_F=2}\rightarrow\ket{F'=3,m_F'=3}\) cycling transition and operated at saturation parameter \(s_0 \approx 5\).
Fluorescence is collected with two high-NA objectives and coupled to multimode fibre arrays (MMFA) for fast, site-resolved photon counting,  which also provides a spatial filter of stray-light background.
To mitigate antitrapping effects and reduce inhomogeneous ac Stark shifts, the tweezers are strobed out of phase with the pulsed probe beams \cite{lukin-bluvstein2024nature, camera-radnaev2025prxquantum}. Before probing, the tweezer depth is raised to \(6~\mathrm{mK}\). 
The fibre-collected photons are detected by APD arrays and processed on the FPGA using a calibrated likelihood-ratio discriminator \cite{supp_mat}.  This photon-counting channel converts fluorescence detection from a fixed camera exposure into a time-resolved measurement record \cite{hume2007high}.

\begin{figure}[ht]
	\centering
	\includegraphics[width=\columnwidth]{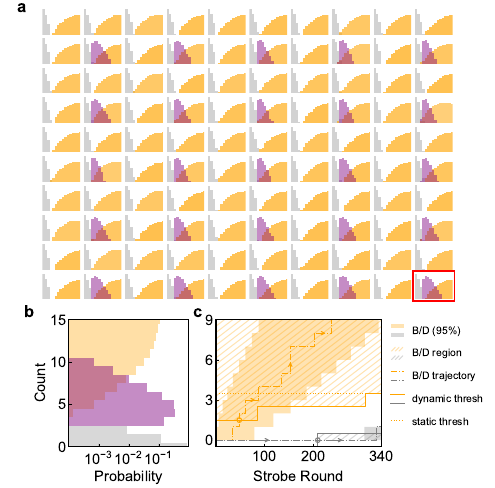}
	\caption{\textbf{Time-resolved photon counting and adaptive strategy.} 
	\textbf{a,} Log-scale APD-count histograms for the full \(10\times10\) atom array. Grey and orange show dark- and bright-state readout, respectively, with mean detected counts of 0.1 and 15. Purple shows bright-state counts on the \(5\times5\) adaptive subarray, for which site-selective hiding truncates the distribution and produces a sub-Poissonian-like profile.
    \textbf{b,} Enlarged histogram for a representative site marked by the red box in \textbf{a}.
    \textbf{c,} Comparison between fixed exposure and dynamic decision. In fixed-exposure readout, the state is assigned from the final integrated count using a single threshold. In dynamic decision, APD clicks are processed during the probe pulse and compared with time-dependent decision boundaries. A bright outcome is identified when crossing a dynamic threshold, and triggers the corresponding hiding beam; a dark outcome can be declared once the lower boundary is crossed. For the representative trajectories shown, the probe duration is reduced to approximately \(1/8\) for bright outcomes and \(1/2\) for dark outcomes. The mean probe time under adaptive readout is \(15~\mu\mathrm{s}\). Each strobing round takes $1~\mu\mathrm{s}$, during which the probe light is on for $0.2~\mu s$. Maximal exposure is capped at $340$ rounds. Shaded areas represent 95\% confidence regions; the dashed areas represent the bright/dark decision regions.
    } 
    \label{fig2}
\end{figure}

Real-time discrimination allows an adaptive imaging strategy that suppresses unnecessary scattering across the array.
We demonstrate this capability on a \(5\times5\) subarray, highlighting a key distinction from conventional camera-based imaging.
A 25-channel \(852~\mathrm{nm}\) hiding system, each gated by a fibre AOM, is delivered to the atoms through a \(5\times5\) polarization-maintaining fibre array (PMFA).
A microlens array (MLA) integrated on the PMFA output facet expands the beam waists, providing the desired waist-to-pitch ratio while maintaining negligible crosstalk between the \(5~\mu\mathrm{m}\) neighbouring sites.
Before reaching the atom array, the hiding beams are set to \(\pi\)-polarization to suppress V-type depumping into the dark state \cite{apd-chow2023physreva}.
During adaptive imaging, once the FPGA identifies a site as bright, it immediately triggers the corresponding channel to turn on the hiding beam.
The atom is thereby decoupled from the probe light, protecting it from further photon scattering and associated recoil heating \cite{apd-gibbons2011physrevlett, apd-chow2023physreva, hu2025sitecavity, camera-lee2026rapidcavity}. The end-to-end feedforward latency is less than \(1~\mathrm{\mu s}\)--- negligible on the scale of the probe duration and orders of magnitude below a camera frame period --- so each atom is shielded essentially the moment its state is known.

Figure~\ref{fig2} (a) shows the APD-count distributions from the full \(10\times10\) atom array on a logarithmic scale.
Bright and dark states yield, on average, 15 and 0.1 detected counts, respectively, as shown by the orange and grey histograms.
With adaptive hiding enabled, the bright-state count distribution, shown in purple, becomes substantially narrower and exhibits a truncated, sub-Poissonian-like profile.
In conventional fixed-exposure imaging, Poissonian broadening of the bright-state distribution requires the exposure time to accommodate rare trajectories near the decision threshold.
Most bright-state trajectories, however, become distinguishable much earlier.
Real-time identification therefore allows those atoms to be shielded before the nominal exposure ends.
Averaged over outcomes and sites, the cumulative probe-on time is only \(15~\mu\mathrm{s}\), compared with \(68~\mu\mathrm{s}\) at the maximum \(340~\mu\mathrm{s}\) readout window. Adaptive protection therefore reduces the mean photon-scattering exposure to approximately one-fifth of its worst-case value.

The same time-resolved record enables dynamic decision thresholds, rather than a single threshold on the final integrated count with conventional detection strategies \cite{camera-kwon2017physrevlett, camera-martinez-dorantes2017physrevlett, camera-wu2019natphys, camera-graham2023physrevx, camera-lis2023physrevx,camera-nikolov2023physrevlett,camera-scott2025physrevlett, camera-radnaev2025prxquantum}.
At each time step, pre-measured bright and dark count distributions define the lower and upper decision boundaries: for a target fidelity of 99.99\%, a bright event is declared once the accumulated count exceeds the corresponding upper quantile of the dark-state distribution, and vice versa, as illustrated in Fig.~\ref{fig2} (c).
At this operating point, the dynamic thresholds reduce the average probe time by about an order of magnitude for bright outcomes and a factor of two for dark outcomes relative to fixed-exposure readout.
More general real-time inference, such as integrated Bayesian analysis \cite{ion-myerson2008physrevlett}, could further improve the speed--fidelity trade-off by identifying atom-loss events and non-Poissonian after-pulse counts in the photon record.

\begin{figure}[b!]
	\centering
	\includegraphics[width=\columnwidth]{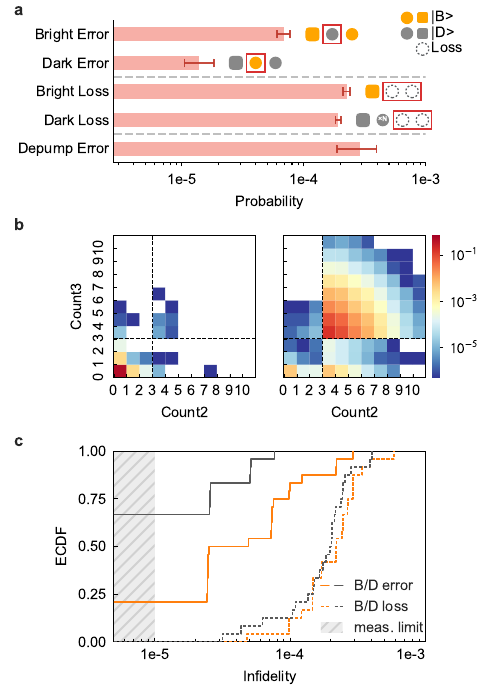}
	\caption{\textbf{Model-free benchmarking of FNDR fidelity and atom loss.}
    Three consecutive FNDR measurements are applied to atoms prepared in either the bright or the dark state, followed by a state-insensitive atom-presence image to identify atom loss.
    \textbf{a,} Detection strings and error decomposition.  The extracted error probabilities, averaged over the \(5\times5\) adaptive subarray, are \(6.9^{+0.9}_{-0.8} \times 10^{-5}\) for bright-state assignment, \(1.4^{+0.4}_{-0.3} \times 10^{-5}\) for dark-state assignment, \(2.2^{+0.2}_{-0.1}\times10^{-4}\) for bright loss, and \(1.9^{+0.1}_{-0.1}\times10^{-4}\) for dark loss. Squares denote preparation state, solid circles denote FNDR outcomes, and dashed circles denote loss in atom-presence image. Red boxes mark first-order error strings.  Normalized string occurrences are plotted on a logarithmic scale.
    \textbf{b,} Two-measurement quadrant histogram for the last two FNDR results. Left panel -- dark state preparation: the bottom-left quadrant corresponds to correct successive dark outcomes. The bottom-right and top-left quadrants indicate dark-state assignment errors, and the top-right quadrant indicates state-preparation error. Right panel -- bright state preparation: the top-left quadrant indicates a bright-state discrimination error, the bottom-left represents a state-preparation error, and the bottom-right quadrant encompasses depumping (predominantly the hiding assisted process, which occurs after state identification and causes no readout error; see main text), atom loss, and bright-state discrimination errors. 
    \textbf{c,} Empirical cumulative distribution function (ECDF) of per-channel base-discrimination error across the 25 adaptive readout channels.
    }
	\label{fig3}
\end{figure}

\begin{figure*}[ht]
	\centering
    \includegraphics[width=\textwidth]{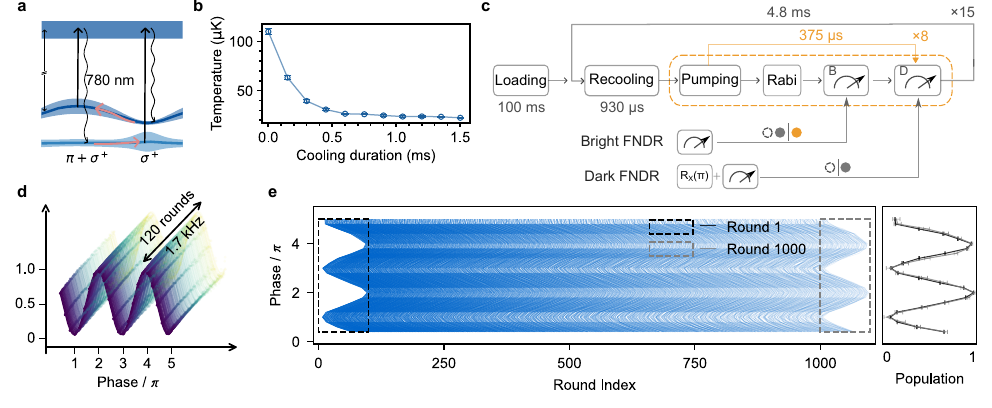}
	\caption{\textbf{Fast FNDR enables high-clock-rate repeated circuits with atom reuse.} 
    Repeated Raman-driven \cite{levine2022dispersive} Rabi circuits are operated with loss-resolved FNDR readout, reaching 120 rounds at a \(1.7~\mathrm{kHz}\) clock rate, and more than \(10^3\) rounds at \(0.7~\mathrm{kHz}\) with per-round recooling.  
    \textbf{a,} Finite-field grey-molasses cooling compatible with the \(10~\mathrm{G}\) bias field during readout. A representative
    one-dimensional cut of the moving $\pi/\sigma^+$ dressed-state landscape based on simulation \cite{supp_mat} is shown; line thickness indicates the local photon-scattering rate. 
    The simulation shows that the Sisyphus mechanism underlying grey-molasses cooling remains efficient at finite magnetic field, provided that a sufficient polarization gradient is maintained.
    \textbf{b,} Temperature of trapped atoms versus grey-molasses cooling time, extracted from Monte Carlo fits to release-and-recapture measurements \cite{tweezer-tuchendler2008physreva}.
    \textbf{c,} Pulse sequence for the repeated Raman-driven Rabi circuit operated at a \(1.7~\mathrm{kHz}\) clock rate. Two FNDR measurements separated by an \(R_x(\pi)\) pulse distinguish \(\ket{1}\), \(\ket{0}\), and atom loss through the outcome pairs \((B,D)\), \((D,B)\), and \((D,D)\), respectively. Because FNDR injects little heating, recooling is applied only once every eight rounds.  
    \textbf{d,} Measured results of the 1.7 kHz high-clock-rate repeated Rabi circuit.
    \textbf{e,} Operation over \(10^3\) rounds at a \(0.7~\mathrm{kHz}\) clock rate with recooling after each round. Dashed boxes indicate the population oscillation scale, and the right inset compares the first and last rounds. Error bars denote 95\% confidence intervals.
    } 
	\label{fig4}
\end{figure*}

\vspace{1em}
\noindent \textbf{Model-free readout error benchmarking}

\noindent We distinguish three readout error channels: base bright--dark discrimination error, measurement-induced atom loss, and hyperfine depumping from the bright to the dark manifold. The first is an incorrect fluorescence assignment conditioned on the atom remaining in the intended manifold \cite{huie2023repetitive, camera-radnaev2025prxquantum, camera-scott2025physrevlett,yb-muzi2025microsecond}. Loss is identified by a final state-insensitive presence image. Depumping changes the internal state while the atom remains trapped; it contributes to the readout error only if it occurs before the state decision is made.

We benchmark the base discrimination infidelity and loss using a model-free sequence of three consecutive FNDR measurements followed by atom presence measurement \cite{norcia2024iterative, largearray-manetsch2025nature}.
Unlike model-derived benchmarks, which rely on assumptions about the underlying photon count profile, the model-free protocol determines error rates directly from measured event strings.
After initially preparing the atoms in either the bright or dark state, the first FNDR measurement verifies state preparation with high confidence.
Conditioned on this verification, the next two FNDR measurements identify first-order error strings, while the final presence image separates trapped atoms from loss events. Because all error rates are small, higher-order strings are negligible at the quoted precision.

Representative event strings are shown in Fig.~\ref{fig3} (a).
For example, an atom prepared and verified in the bright state that gives the string 101 corresponds to a false dark result during the second FNDR measurement, followed by recovery to the correct bright outcome in the third measurement.
From the corresponding string probabilities, we extract bright- and dark-state discrimination error rates of \(6.9^{+0.9}_{-0.8} \times 10^{-5}\) and \(1.4^{+0.4}_{-0.3} \times 10^{-5}\), respectively, giving a state-averaged base discrimination infidelity of \(4.1^{+0.7}_{-0.6} \times 10^{-5}\). The same protocol yields
bright- and dark-state atom-loss probabilities of \(2.2^{+0.2}_{-0.1}\times10^{-4}\) and \(1.9^{+0.1}_{-0.1}\times10^{-4}\), corresponding to a state-averaged loss of \(2.1^{+0.1}_{-0.1}\times10^{-4}\).
A detailed analysis of the error strings based on a Markov state tree is given in \cite{supp_mat}.

We independently quantify depumping error by inserting a controlled probe-light interval into the sequence and identifying the resulting error strings while postselecting on final atom survival \cite{supp_mat}.  This gives a depumping probability of \(2.9^{+1.0}_{-1.0}\times10^{-4}\) during the readout, caused by the probe light's residual misalignment with the quantization axis and small polarization impurity. 
There is an additional hiding-beam-assisted depumping pathway possible when the \(852~\mathrm{nm}\) hiding light and probe are simultaneously applied \cite{supp_mat}.
In our protocol, however, the hiding beam is activated only after the atom has been identified as bright, so hiding-beam-assisted depumping does not contribute to readout error.

\begin{figure}[t]
	\centering
    \includegraphics[width=\columnwidth]{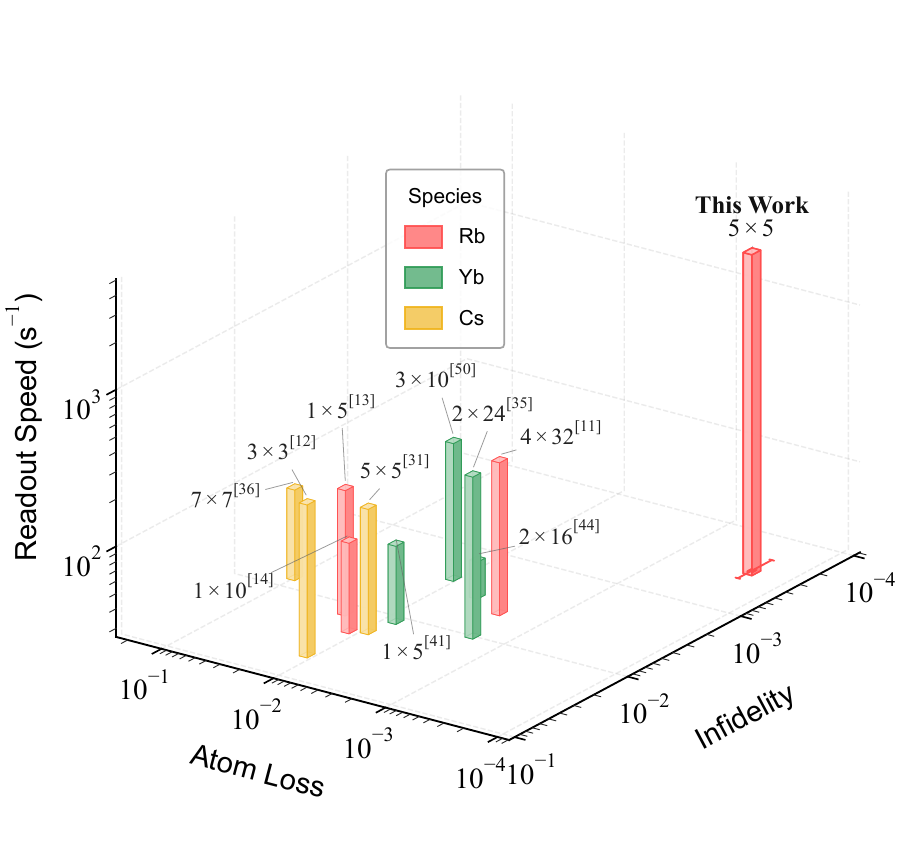}
	\caption{\textbf{Readout performance compared with previous neutral-atom array demonstrations.} 
    Nondestructive state-resolving readout of free-space atom arrays are compared in terms of readout infidelity, atom-loss probability and readout speed \cite{camera-graham2023physrevx,camera-kwon2017physrevlett,camera-martinez-dorantes2017physrevlett,camera-lis2023physrevx,huie2023repetitive,camera-radnaev2025prxquantum,camera-bluvstein2026nature,camera-nikolov2023physrevlett,muniz2025repeated,norcia2023midcircuit}.
    Readout speed is defined as the inverse of the maximal exposure time; for this work we therefore quote the worst-case readout window of \(340~\mu\mathrm{s}\). The adaptive strategy in fact terminates most measurements far earlier, reducing the mean probe-on time per atom to \(15~\mu\mathrm{s}\), so the comparison shown here is conservative.
    Infidelity includes depumping error when independently reported; for the present work this corresponds to the base-discrimination infidelity plus the measured hyperfine-depumping contribution.
    The FNDR reaches the joint high-rate, low-loss and low-infidelity regime, rather than improving one metric at the expense of the others.
    } 
	\label{fig5}
\end{figure}

Taken together, this operating point improves simultaneously on the lowest infidelity and the lowest atom loss previously reported for array-scale qubit readout, by roughly an order of magnitude each \cite{camera-radnaev2025prxquantum,camera-bluvstein2026nature}, while operating an order of magnitude faster than the millisecond-scale exposures of camera-based schemes even for our longest readout window (Fig.~\ref{fig5}).

\vspace{1em}
\noindent \textbf{High-clock-rate circuits with atom reuse}

\noindent The operational value of a nondestructive measurement is set by its performance inside a repeated circuit.
In quantum error correction, ancilla measurements must be performed many times for syndrome extraction to protect the data qubits \cite{terhal2015quantummemory}.
The speed and reusability of these measurement cycles set the clock of fault-tolerant computation \cite{google2025belowthreshold,camera-bluvstein2026nature,muniz2025repeated,atomcomputing2026toricQEC,gidney2025lessthanmillion}.
Resource estimations for neutral-atom processors commonly assume a \(1~\mathrm{ms}\) clock period \cite{1ms-zhou2025resource, 1ms-cain2026shor, 1ms-ismail2026star}.
Experiments with atom reuse have so far been slower, typically requiring tens of milliseconds per round \cite{muniz2025repeated, norcia2023midcircuit, camera-graham2023physrevx, huie2023repetitive, camera-radnaev2025prxquantum, continuous-li2025fast}, with the fastest previous demonstration at \(4~\mathrm{ms}\) \cite{camera-bluvstein2026nature}. This gap arises largely from slow and lossy readout.

FNDR reduces both irreversible loss and accumulated motional excitation. Its low photon budget suppresses measurement-induced atom loss and lowers the cooling duty cycle.
Residual heating is removed with grey-molasses cooling \cite{graymolasses-rosi2018scirep}, extended here in a \(\pi/\sigma^+\) configuration that is compatible with the \(10~\mathrm{G}\) field used for readout and Raman control \cite{supp_mat}. 
This pulse resets the atoms to \(\sim20~\mu\mathrm{K}\) in \(800~\mu\mathrm{s}\) (Fig.~\ref{fig4} a,b), avoiding the need to switch to a zero-field cooling configuration.

For the repeated-circuit demonstrations, we encode the qubit in the stretched states
\(\ket{0}\equiv\ket{F=1,m_F=1}\) and
\(\ket{1}\equiv\ket{F=2,m_F=2}\), which are dark and bright under FNDR,
respectively.
Figure~\ref{fig4} (c-d) shows a representative 120-round Rabi circuit operated at a \(1.7~\mathrm{kHz}\) repetition rate.
Because the FNDR photon budget is small, recooling is required only once every eight rounds while maintaining high overall survival.

Applying the same finite-field recooling pulse after every round increases the survival over longer circuits, allowing operation beyond \(10^3\) rounds at a 0.7-kHz clock rate (Fig.~\ref{fig4} e). 
At these clock rates, more than thousands of measurement–reset cycles fit within the second-scale coherence times of current tweezer arrays \cite{largearray-manetsch2025nature}, so mid-circuit measurement no longer dominates the coherence budget available for syndrome extraction; the remaining coherence limits are then addressed at the logical level by repeated error correction \cite{atomcomputing2026toricQEC}.

\vspace{1em}
\noindent \textbf{Outlook}

\noindent We have demonstrated a fast nondestructive readout architecture that removes measurement as the clock-rate bottleneck of neutral-atom quantum computing, bringing repeated circuit operation with atom reuse into the sub-millisecond regime while simultaneously advancing the state of the art in readout fidelity and atom survival (Fig.~\ref{fig5}). 
Beyond raw speed, the site-parallel classification and sub-microsecond feedforward loop eliminate the frame latency inherent to camera-based readout, providing a low-latency readout interface demanded by real-time syndrome decoding and high-throughput, measurement-conditioned logical operations \cite{fowler2012realtimedecoding, caune2026realtimedecoding, gottesman1999mbqc, litinski2019surgery}.
The FNDR principle is not tied to the particular alkali implementation used here; it can be combined with other species and encodings, including alkaline-earth-like atoms with metastable manifolds for leakage or erasure detection \cite{wu2022erasure, erasure-ma2023nature,scholl2023erasure, chow2024leakage, senoo2025highfidelitycoherent}.

Technical improvements should allow the circuit repetition rates to approach \(\sim5~\mathrm{kHz}\) \cite{supp_mat}, which can be further improved with higher-NA photon collection. 
The near-term clock constraints of the neutral-atom quantum processor now shift from measurement to fast atom transport and mid-circuit cooling \cite{saffman2025connectivitycost, tsai2026gate}. Scaling in channel number is primarily an engineering rather than architectural question: the current \(10\times10\) detection channels and \(5\times5\) adaptive-hiding channels are constrained only by the APDs and optical switches available in our laboratory, not by the optical layout.
Extending to thousands of sites will require more densely integrated APD arrays \cite{morimoto2020megapixelapd, wang2023apdarray} and electro-optic switch or modulator arrays \cite{christen2025integratedswitch,zhao2025integratedswitch}. The fibre--microlens delivery system could be replaced by monolithic three-dimensional waveguide arrays that perform pitch conversion and channel routing in a passive glass chip \cite{gattass2008femtosecond, wang2024laserdirectwriting,maeda2025glasswaveguide}.

The immediate application is repeated mid-circuit measurement for quantum error correction \cite{dualspecies-singh2023mid,lin2026sustain,norcia2023midcircuit,camera-lis2023physrevx}.
In a zone-based processor, the FNDR can serve as a high-speed measurement zone: ancillas are measured  with high fidelity and low loss, reset, and returned or replenished within a short time, preventing measurement from setting back the logical cycle \cite{camera-bluvstein2026nature, muniz2025repeated, atomcomputing2026toricQEC}.
In architectures with fast site- or species-selective control and reduced physical shuttling \cite{fiberandcavity-li2025natcommun, camera-radnaev2025prxquantum, lib2026velocityenabled,dumke2002micro,cesa2023universal,anand2024dual, senoo2025highfidelitycoherent} readout window becomes a particularly direct constraint: the measurement duration directly sets the interval over which data qubits must remain shielded or shelved. The FNDR can therefore substantially reduce dephasing, photon-scattering errors and control errors accumulated during these intervals \cite{camera-graham2023physrevx,camera-lis2023physrevx,ion-crain2019highspeed}.
Across these architectures, compressing the measurement-and-feedforward stage increases the number of syndrome-extraction rounds and measurement-conditioned logical operations that can be completed per unit wall-clock time, while reducing idling errors accumulated in each round.
Combined with higher-NA collection or cavity-compatible interfaces \cite{fiberandcavity-shaw2026nature, fiberandcavity-li2025natphys}, such an integrated measurement layer would connect the temporal efficiency shown here to the space--time scaling demanded by large-scale and distributed architectures \cite{ramette2024linkfault}.
Measurement would then act not as the rate-limiting step of a neutral-atom processor, but as a fast, parallel and reusable resource, paving the way for high-throughput, fault-tolerant quantum computation.

\begin{acknowledgments}
\textbf{Acknowledgments}

This work was supported by the National Natural Science Foundation of China (No. 12322415), Quantum Science and Technology-National Science and Technology Major Project (2021ZD0301405), the HFNL Self-Deployed Project (ZB2024010101, ZB2024010102, ZB2024010201, and ZB2024010501), the Shanghai Municipal Science and Technology Commission (24DP2600300) and the New Cornerstone Science Foundation.
\end{acknowledgments}


We thank Xiang Zhang, Junhua Zhang, Ye Wang, Yong-Heng Huo for helpful discussions and assistance.

\bibliographystyle{apsrev4-2}
\bibliography{ndr}

\clearpage

\end{document}